\documentclass[12pt]{article}
\usepackage{graphicx}
\usepackage{amsmath}

\def\Eq#1{Eq.~(\ref{#1})}

\newcommand\pubnumber{LPN14-125, TTP14-032, IFIC/14-75}
\newcommand\pubdate{November 15, 2014}

\def\ific{Instituto de F\'{\i}sica Corpuscular, Universitat de Val\`encia -
Consejo Superior de Investigaciones Cient\`{\i}ficas, Parc Cient\'{\i}fic, 
E-46980 Paterna, Valencia, SPAIN}
\def\KA{Institut f\"ur Theoretische Teilchenphysik, 
Karlsruher Institut f\"ur Technologie, D-76133, Karlsruhe, GERMANY}

\def\support{\footnote{Work supported by the 
Research Executive Agency (REA) of the European Union under 
the Grant Agreement number PITN-GA-2010-264564 (LHCPhenoNet),
by the Spanish Government and EU ERDF funds 
(grants FPA2011-23778 and CSD2007-00042 
Consolider Project CPAN) and by GV (PROMETEUII/2013/007).}}

\def\Title#1{\begin{center} {\Large #1 } \end{center}}
\def\Author#1{\begin{center}{ \sc #1} \end{center}}
\def\Address#1{\begin{center}{ \it #1} \end{center}}

\newcommand\pubblock{\rightline{\begin{tabular}{l} \pubnumber\\
         \pubdate  \end{tabular}}}
\newenvironment{Abstract}{\begin{quotation}  }{\end{quotation}}
\newenvironment{Presented}{\begin{quotation} \begin{center} 
             PRESENTED AT\end{center}\bigskip 
      \begin{center}\begin{large}}{\end{large}\end{center} \end{quotation}}

\begin{document}
\begin{titlepage}
\pubblock

\vfill
\Title{Forward-backward and charge asymmetries \\ at Tevatron 
and the LHC \support}
\vfill
\Author{Johann H. K\"uhn}
\Address{\KA}
\Author{Germ\'an Rodrigo}
\Address{\ific}
\vfill
\begin{Abstract}
We provide a qualitative and quantitative unified picture of the charge 
asymmetry in top quark pair production at hadron colliders in the SM 
and summarise the most recent experimental measurements.  
\end{Abstract}
\vfill
\begin{Presented}
8th International Workshop on the CKM Unitarity Triangle (CKM 2014), 
Vienna, Austria, September 8-12, 2014
\end{Presented}
\vfill
\end{titlepage}
\def\thefootnote{\fnsymbol{footnote}}
\setcounter{footnote}{0}

\section{Introduction}

An interesting property in top quark pair production in hadronic collisions
is the charge asymmetry, namely a difference in the angular 
distribution of the top quarks with respect to that of the antiquarks, 
due to higher order corrections in the Standard Model (SM). 
Since 2007, sizeable differences have been observed between theory 
predictions~\cite{Kuhn:1998kw,Bowen:2005ap,Antunano:2007da} 
and measurements 
by the CDF~\cite{Weinelt:2006mh,Hirschbuehl:2005bj,Schwarz:2006ud,Aaltonen:2008hc,Aaltonen:2011kc,Aaltonen:2012it}
and the D0~\cite{Abazov:2007ab,Abazov:2011rq,Abazov:2014cca} 
collaborations at the Tevatron. 
This discrepancy was particularly pronounced for the subsample of
$t\bar t$ pairs with large invariant mass, $m_{t\bar t} > 450$~GeV,
and the asymmetry defined in the $t\bar t$ rest-frame, 
where a $3\sigma$ effect was advocated~\cite{Aaltonen:2011kc}. 
These anomalies triggered a large number of theoretical 
investigations speculating about possible new physics
contributions~\cite{Antunano:2007da,Ferrario:2009bz,Frampton:2009rk,Djouadi:2009nb,Jung:2009jz,Cheung:2009ch,Shu:2009xf}.
Recent analysis, however, lower this discrepancy, particularly at 
D0~\cite{Abazov:2014cca}. Also, measurements 
at the LHC~\cite{Chatrchyan:2011hk,Chatrchyan:2012cxa,CMS:2013nfa,Chatrchyan:2014yta,ATLAS:2012an,Aad:2013cea,ATLAS:2012sla} 
are in good agreement with the SM prediction. 

The $t\bar t$ asymmetry is often called forward--backward asymmetry 
at the Tevatron and charge asymmetry at the LHC, but in fact, although the 
kinematical configurations of the two machines are different the physical 
origin of the asymmetry in both cases is the same. 
In this talk, we provide a qualitative and quantitative
unified picture of this property in the SM 
and summarize the experimental measurements.

\section{The charge asymmetry in the SM}

The dominant contribution to the charge asymmetry
originates from $q\bar{q}$ annihilation~\cite{Kuhn:1998kw} 
due to the interference between the 
Born amplitudes for $q\bar{q}\to t\bar{t}$ and the one-loop 
amplitudes, which are antisymmetric under the exchange of the heavy quark 
and antiquark (box and crossed box). 
To compensate the infrared divergences, these virtual corrections 
are combined with the interference between initial and 
final state radiation.
Diagrams with the triple gluon coupling in both real and virtual 
corrections give rise to symmetric amplitudes and can be ignored. 
A second contribution to the asymmetry from 
quark-gluon scattering (``flavour excitation'') hardly contributes to 
the asymmetry at the Tevatron. At the LHC, it enhances 
the asymmetry in suitable chosen kinematical regions~\cite{Kuhn:1998kw}. 
CP violation arising from electric or chromoelectric dipole moments 
of the top quark do not contribute to the asymmetry.


The inclusive charge asymmetry is proportional to the symmetric colour 
factor $d_{abc}^2=40/3$, and positive, namely the 
top quarks are preferentially emitted in the direction of the incoming 
quarks at the partonic level~\cite{Kuhn:1998kw}. 
The colour factor can be understood from 
the different behaviour under charge conjugation of the scattering amplitudes 
with the top and antitop quark pair in a colour singlet or colour octet state.  
The positivity of the inclusive asymmetry is a consequence of the fact that 
the system will be less perturbed, and will require less energy, 
if the outgoing colour field flows in the same direction as the incoming 
colour field. On the contrary, the asymmetry of the $t\bar t$+jet sample 
is negative because radiation of gluons requires to decelerate 
the colour charges.   

At Tevatron, the charge asymmetry is equivalent to a forward--backward
asymmetry as a consequence of charge conjugation symmetry, and
arises from the collision of valence quarks and antiquarks of 
similar momenta. Thus, top quarks are preferentially emitted 
in the direction of the incoming protons. 
The LHC is a proton-proton symmetric machine and obviously a forward--backward 
asymmetry vanishes, however, the same charge asymmetry as defined at the 
Tevatron arises from the small $t\bar t$ sample produced by 
annihilation of valence quarks with sea 
antiquarks~\cite{Kuhn:1998kw,Antunano:2007da}. 
Figure~\ref{fig:rap} shows a qualitatively and not to scale
picture of the rapidity distributions of the top and the antitop quarks 
at the Tevatron (left) and the LHC (centre, right). 
Since valence quarks carry on average more momentum than sea antiquarks,
production of top quarks with larger rapidities is preferred in the SM, 
and antitop quarks are produced more frequently at smaller rapidities.

Mixed QED-QCD and EW-QCD corrections~\cite{Kuhn:1998kw} enhance the QCD asymmetry 
by about twenty percent at the Tevatron~\cite{arXiv:1109.6830,Hollik:2011ps}, 
and by $0.13$ at the LHC~\cite{arXiv:1109.6830}. The difference 
is due to the fact contrary to QCD, the QED and EW corrections depend 
on the flavour of the incoming quarks, being the flavour asymmetries
of opposite sign for up and down quarks. 
While the relative importance of $u\bar u$ 
versus $d\bar d$ annihilation is $4:1$ at the Tevatron, 
it is $2:1$ at the LHC. This leads to an small decorrelation in the SM,
that can be exploited to explain the observed discrepancies at the 
Tevatron with respect to the LHC in some beyond the SM 
scenarios~\cite{Drobnak:2012cz}.

\section{SM predictions at the Tevatron and the LHC}

The charge asymmetry at the Tevatron (aka forward--backward asymmetry)
in the laboratory frame is given by either of the following definitions:
\begin{equation}
A_{\rm lab}=\frac{N(y_t>0)-N(y_t<0)}{N(y_t>0)+N(y_t<0)}
= \frac{N(y_t>0)-N(y_{\bar t}>0)}{N(y_t>0)+N(y_{\bar t}>0)} = 0.056 (7)~, 
\label{eq:Alab}
\end{equation}
requiring to measure the rapidity of either $t$ or 
$\bar t$ for each event. Equivalently, the charge asymmetry 
can be defined in the $t\bar t$ rest-frame though the variable
$\Delta y = y_t-y_{\bar t}$:
\begin{equation}
A_{t\bar t}=\frac{N(\Delta y > 0)-N(\Delta y < 0)}
{N(\Delta y > 0)+N(\Delta y < 0)} = 0.087 (10)~,
\label{eq:Attbar}
\end{equation}
which requires to determine both rapidities simultaneously. 
It is important to stress that although $\Delta y$ is invariant 
under boosts, the size of the asymmetry changes from one frame 
to another. Systematics are also different. 
The difference between the SM predictions in \Eq{eq:Alab} and 
\Eq{eq:Attbar} is not due to any improvement of the theoretical calculations,
but $A_{\rm lab} < A_{t\bar t}$ ($A_{\rm FB}$ in the literature)
due to the fact that the boost into the laboratory frame
partially washes out the partonic asymmetry~\cite{Antunano:2007da}. 

\begin{figure}[t]
\begin{center}
\includegraphics[width=4.9cm]{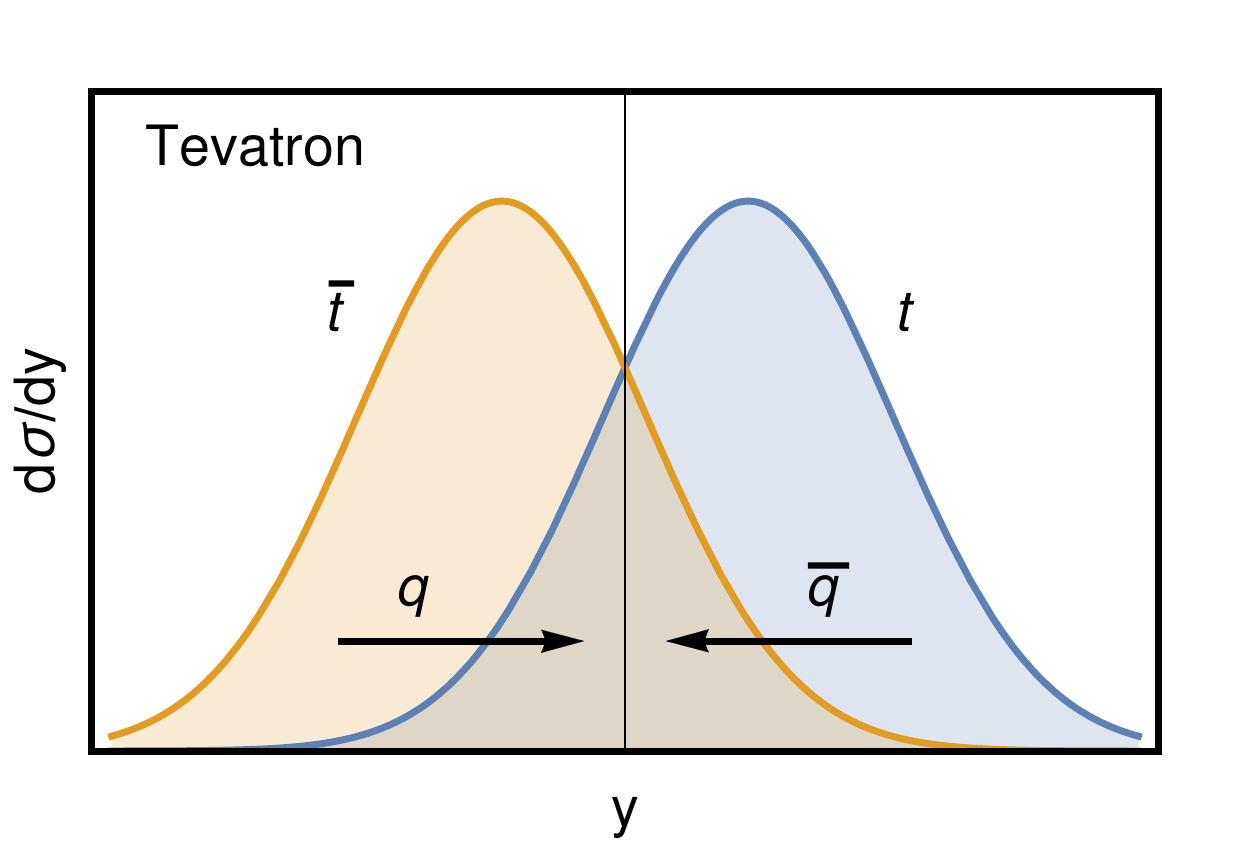} 
\includegraphics[width=4.9cm]{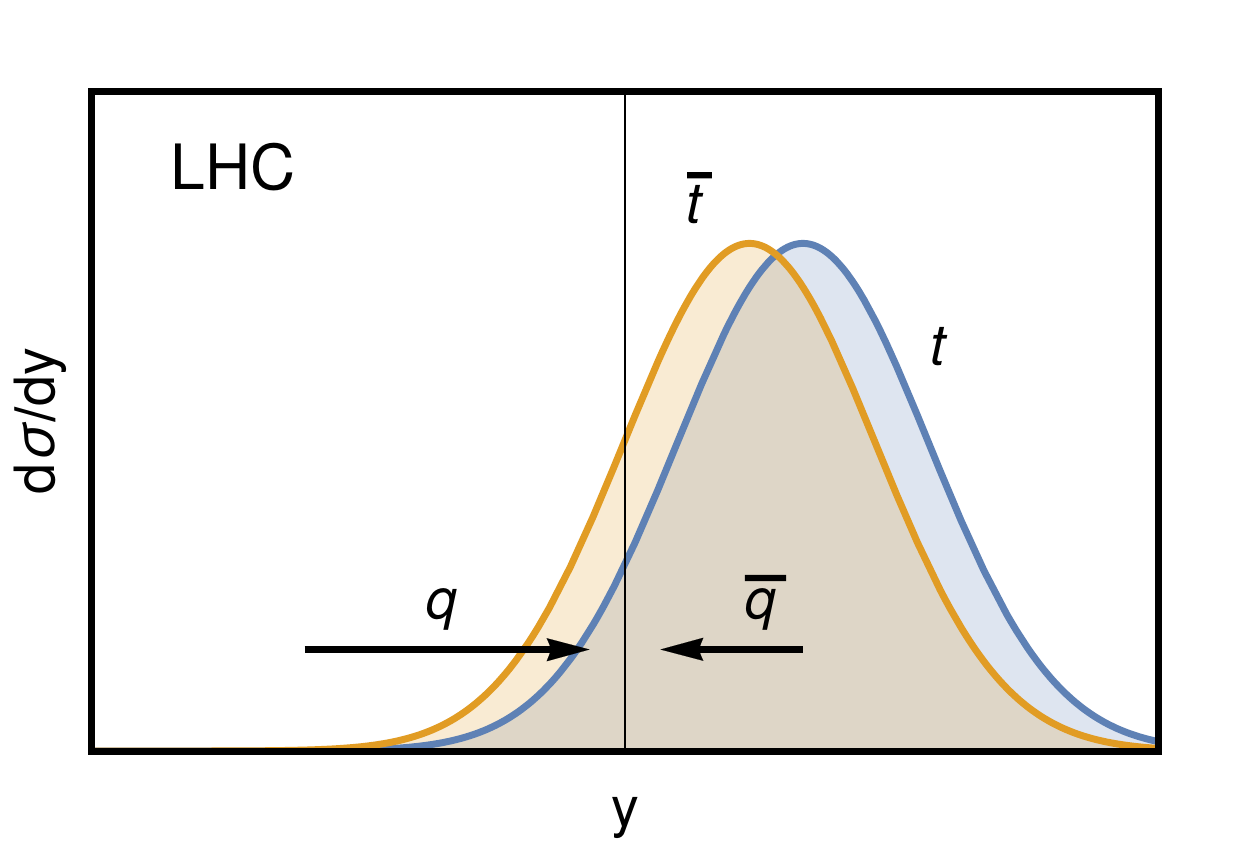}
\includegraphics[width=4.9cm]{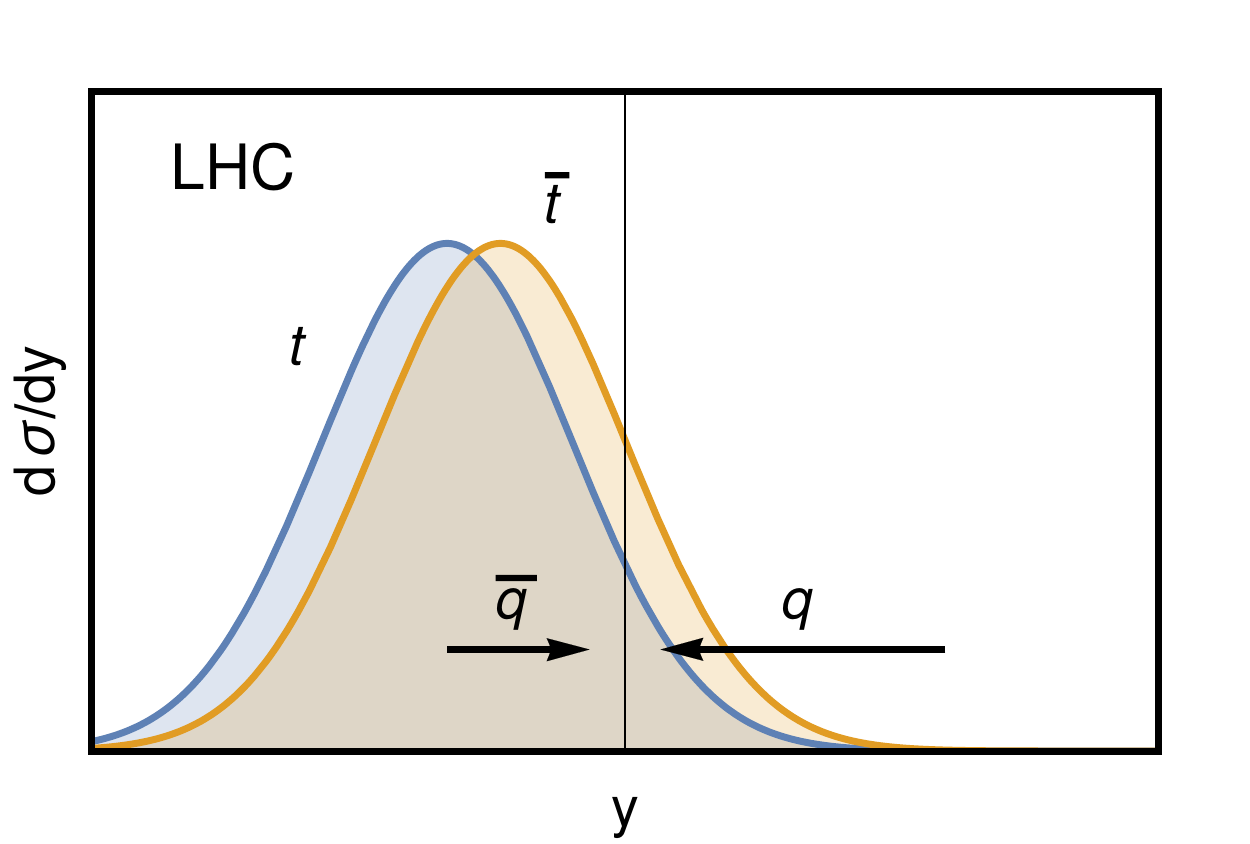}
\caption{Not to scale partonic rapidity distributions of top and antitop quarks 
at the Tevatron (left) and the LHC (centre, right).
\label{fig:rap}}
\end{center}
\end{figure}

At the LHC, the charge asymmetry is defined
through $\Delta |y| = |y_t|-|y_{\bar t}|$:
\begin{equation}
A_C = \frac{N(\Delta |y|>0)-N(\Delta |y|<0)}
{N(\Delta |y|>0)+N(\Delta |y|<0)}
= \begin{cases}
0.0115 (6) @ 7 {\rm TeV} \\
0.0102 (5) @ 8 {\rm TeV} \\
0.0059 (3) @ 14 {\rm TeV} 
\end{cases}~.
\label{AC}
\end{equation}
$\Delta |y|$ is positive (negative) if the product $(y_t+y_{\bar t}) \Delta y$ 
is positive (negative). The factor $Y_{t\bar t} = (y_t+y_{\bar t})/2$,
is the average rapidity of the $t\bar t$ system, and determines whether 
the event is mostly forward ($Y_{t\bar t} > 0$) or backward ($Y_{t\bar t} < 0$), 
and $\Delta y$ is the same variable which is used to measure the 
asymmetry at the Tevatron (see again Fig.~\ref{fig:rap}).

At the LHC, $t\bar t$ production, contrary to 
what happens at the Tevatron, is dominated by gluon fusion 
which is symmetric. Also, the asymmetry at the LHC decreases 
at higher energies because of the larger gluon fusion contribution. 
Therefore, in order to reach a sizeable asymmetry at the LHC
it is necessary to introduce selection cuts to suppress as much as possible 
the contribution of gluon fusion events, and to enrich the sample 
with $q\bar q$ events. In particular, gluon fusion is dominant in the central 
region and can be suppressed by introducing a cut in the average rapidity $Y_{t\bar t}$
(or selecting events with large $m_{t\bar t}$).
Obviously this is done at the price of lowering the statistics, which, 
however, will not be a problem at the LHC at long term. 
A similar asymmetry effect is expected in bottom quark 
production, although it is affected by a higher gluon
fusion dilution~\cite{Kuhn:1998kw,Aaij:2014ywa}, 
even at the Tevatron~\cite{Abazov:2014ysa}.

\begin{figure}[t]
\begin{center}
\includegraphics[width=8cm]{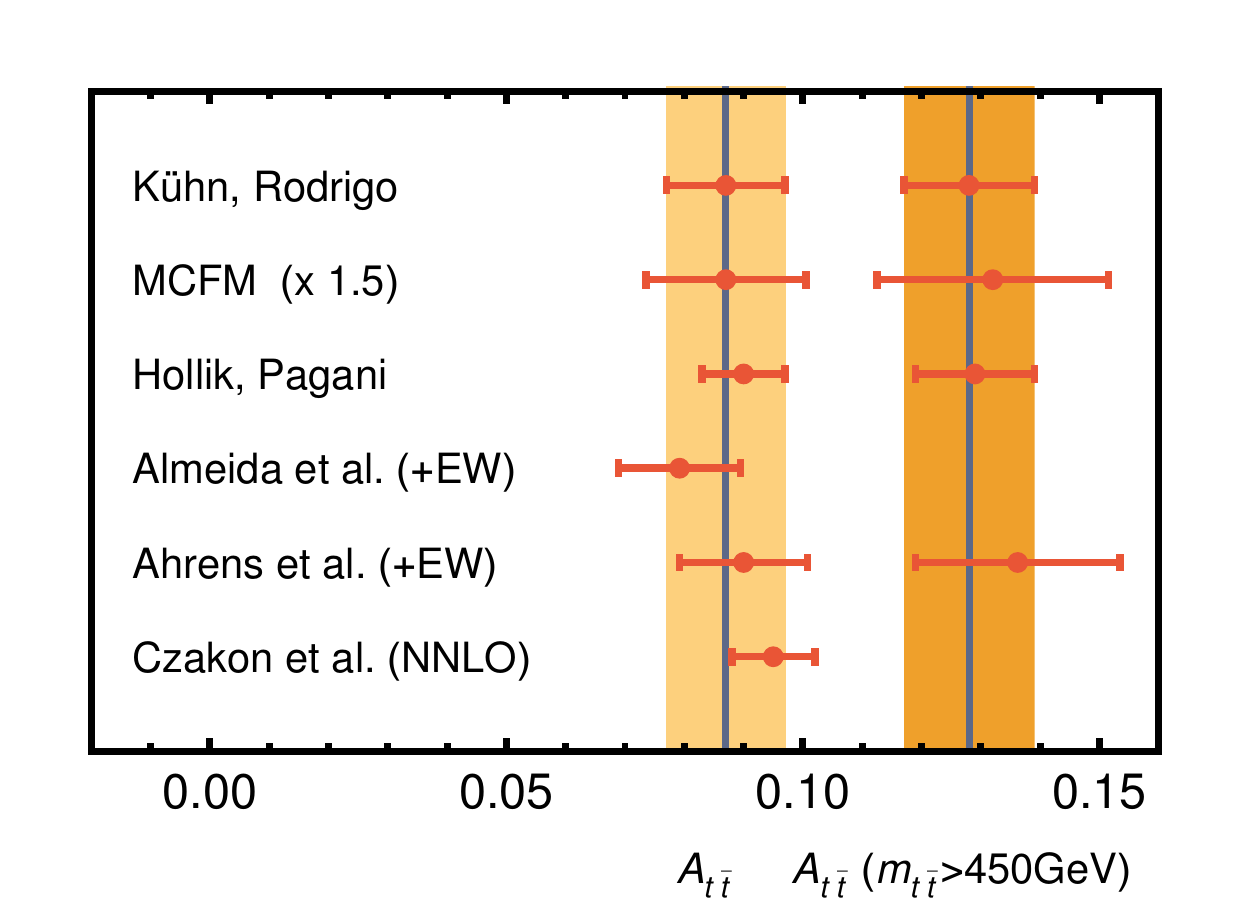}
\caption{Summary of theoretical predictions for the inclusive charge asymmetry 
at the Tevatron in the $t\bar t$ rest-frame, $A_{t\bar t}$, and in the large 
invariant mass region $A_{t\bar t}(m_{t\bar t}>450$~GeV$)$.
\label{fig:th}}
\end{center}
\end{figure}

\begin{figure}[t]
\begin{center}
\includegraphics[width=7cm]{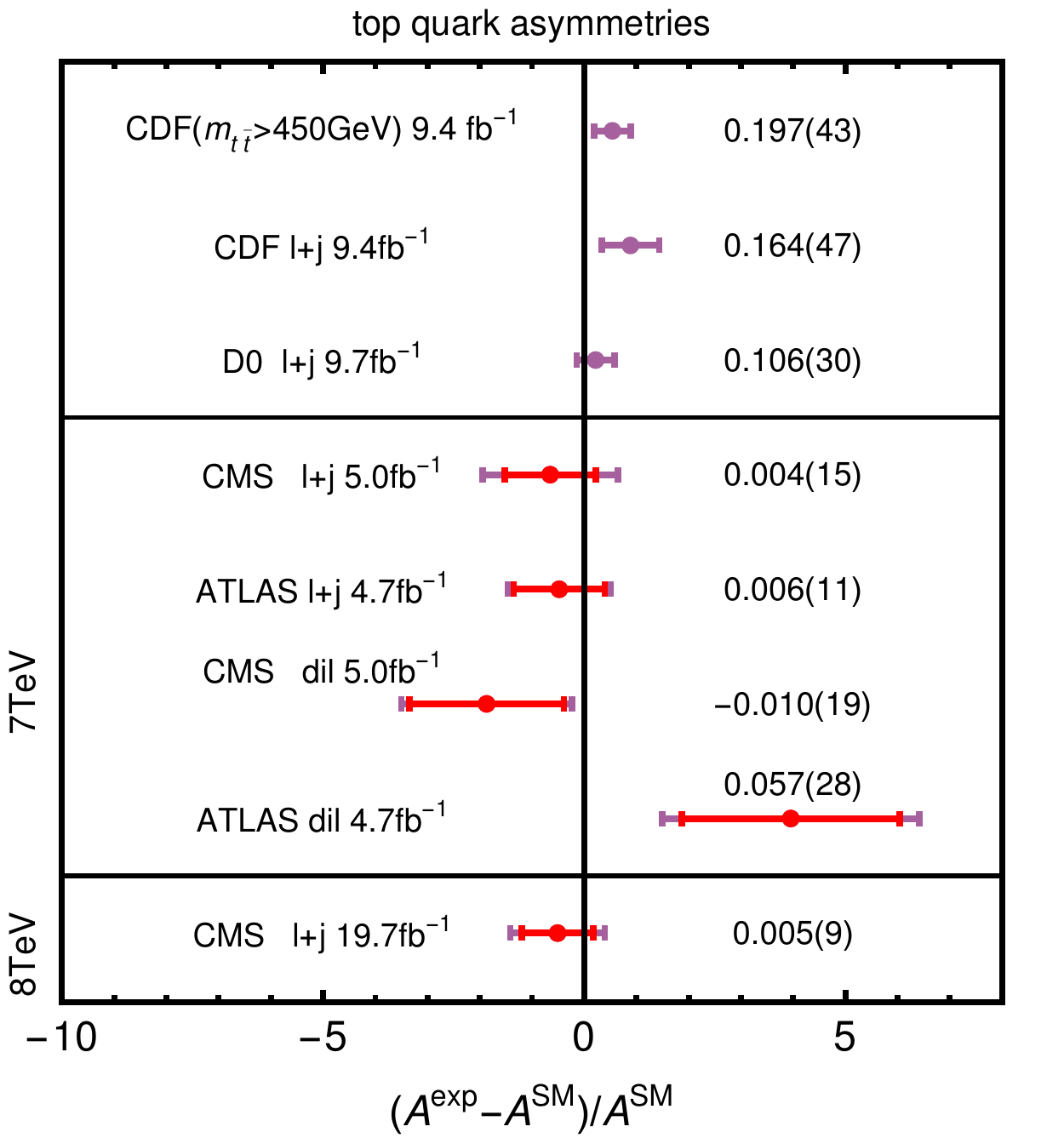}
\includegraphics[width=7cm]{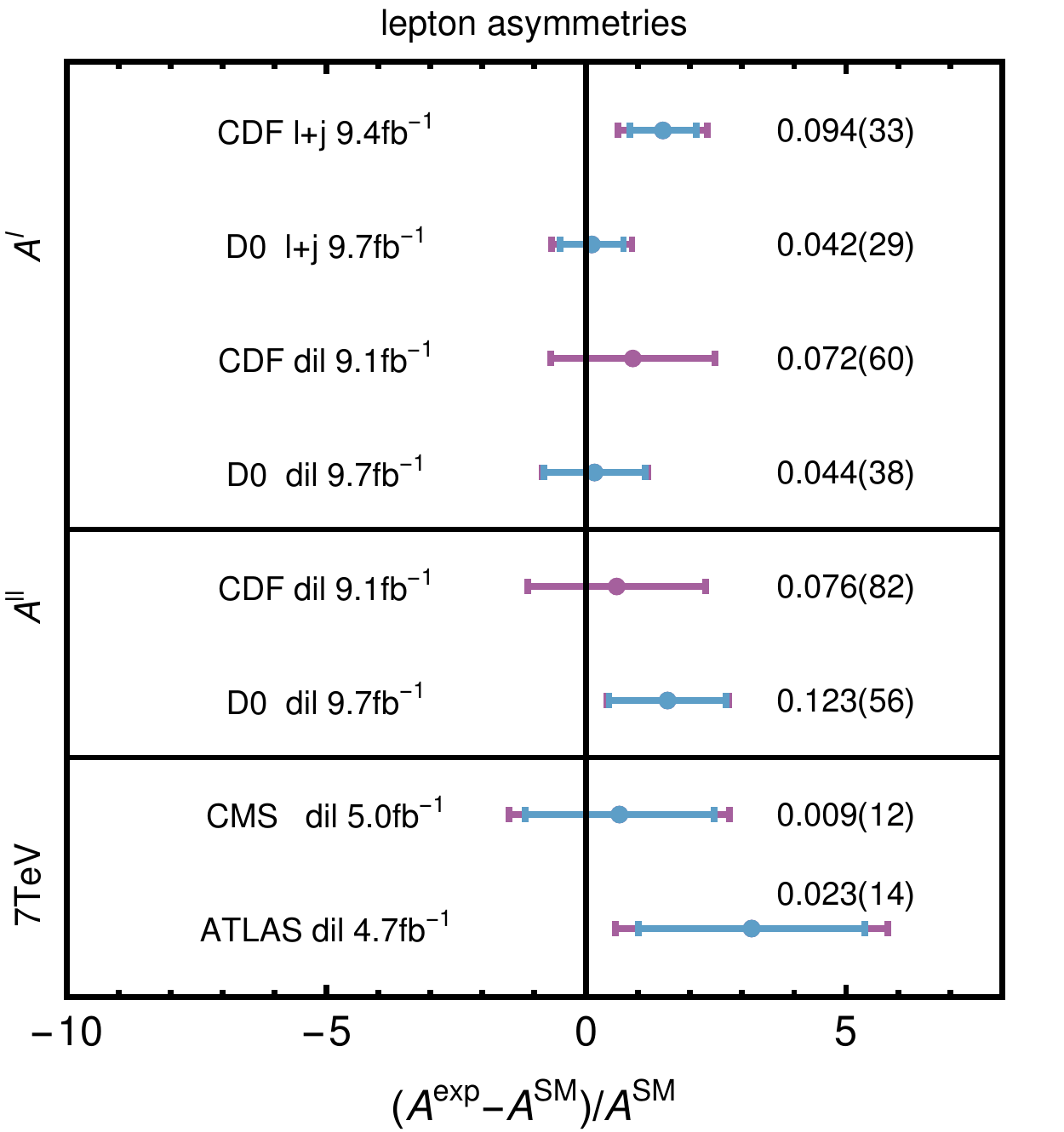}
\caption{Summary of experimental measurements for the top quark and lepton 
asymmetries in the Tevatron and the LHC in comparison with the corresponding
theoretical predictions. 
\label{fig:th2}}
\end{center}
\end{figure}

The charge asymmetry is the ratio of the antisymmetric cross-section 
to the symmetric cross-section. The leading order contribution to the 
antisymmetric cross-section is a loop effect,
but the leading order contribution to the symmetric cross-section 
appears at the tree-level. This suggest that the charge asymmetry 
should be normalised to the Born cross-section~\cite{Kuhn:1998kw}, 
and not the NLO cross-section, in spite of the fact that 
the later is well known, and is included in several Monte Carlo event 
generators such as MCFM~\cite{Campbell:1999ah}.
This procedure is furthermore supported by the fact that theoretical 
predictions resuming leading logarithms (NLL~\cite{Almeida:2008ug} 
and NNLL~\cite{Ahrens:2010zv}) do not modify significantly the central 
prediction for the asymmetry, and are less sensitive to the normalisation. 
Also, recent results on the asymmetry at NNLO~\cite{mitov}, $A_{t\bar t} = 0.095(7)$, 
are within the error bar in \Eq{eq:Attbar}, and confirm the robustness 
of the approximation adopted in Ref.~\cite{Kuhn:1998kw}. 

Figure~\ref{fig:th} summarizes the state-of-the-art SM predictions 
for the inclusive asymmetry in the $t\bar t$ rest-frame, and 
in the large invariant mass region, $m_{t\bar t}>450$~GeV, from different 
authors~\cite{arXiv:1109.6830,Aaltonen:2011kc,Hollik:2011ps,Almeida:2008ug,Ahrens:2010zv}. 
In order to have a coherent picture, EW corrections have been added
to the predictions presented in~\cite{Aaltonen:2011kc,Almeida:2008ug,Ahrens:2010zv}, 
which amount to a factor of about $1.2$, and the Monte Carlo based 
prediction has also been corrected by an extra factor of $1.3$ 
to account for the normalisation to the NLO cross-section.
A nice agreement if found among the different theoretical predictions. 
The small differences are only due to the choice of the 
factorisation and renormalisation scales;
the asymmetry is proportional to the strong coupling. 

The asymmetry can be defined also through the decay products in the
dilepton and lepton+jets
channels~\cite{Chatrchyan:2014yta,ATLAS:2012sla,Aaltonen:2013vaf,Aaltonen:2014eva,Abazov:2012oxa,Abazov:2013wxa,Abazov:2014oea}.
The direction of the lepton (antilepton) is correlated with the direction
of the top quark (top antiquark), particularly for very boosted tops.
The same asymmetries as in \Eq{eq:Alab} to \Eq{AC} can be used 
with the substitutions $y_t \to y_\ell$, $\Delta y \to \Delta y_\ell$, and
$\Delta |y| \to \Delta |y_\ell|$.
Leptons are well measured experimentally, however the asymmetries are diluted
by roughly a factor two~\cite{Bernreuther:2012sx}, at least in the
SM where the top quarks are produced almost unpolarised.
BSM contributions might polarise the top quarks,
then altering the correlation of the top asymmetries with
the lepton asymmetries and spin correlations in BSM scenarios.

A summary of the most recent experimental measurements
in comparison with the respective theoretical predictions
in the SM is presented in Fig.~\ref{fig:th2} (left) for the
top quark asymmetries, and in Fig.~\ref{fig:th2} (right) for the
lepton asymmetries. A good agreement is found with the SM 
with the exception of very few mild discrepancies.

\section{Summary}

The most recent measurements of the top quark asymmetries at
the Tevatron are closer to the SM, although a few mild anomalies
still persist which cannot unfortunately be clarified with further data.
The agreement is, however, not due to relevant enhancements of
the SM predictions. The theoretical predictions have not changed
significantly since the pioneering works, if the correct frame is
chosen for comparison with data; the bulk of the QED and EW
corrections were already included in Ref.~\cite{Kuhn:1998kw} and the
recent reevaluations increase the central value by only +0.008.
Very recent NNLO results lie within the previously quoted theoretical
error band and confirm the appropriateness of the
long discussed question about the normalisation of the asymmetries.
Although the current measurements leave a very small window for BSM,
the existence of these anomalies since 2007 have clearly boosted a
better understanding of the properties of the top quark,
both for model building and precision physics.
Plenty of room for further analysis of the top quark, lepton
and bottom quark asymmetries at the LHC exists. In particular, asymmetries 
are sensitive to BSM and still complementary to other observables for 
BSM searches.

\end{document}